\documentstyle[12pt,psfig,axodraw,a4]{article} 
\def\bild#1#2{    
        \vspace*{-5mm}
        \begin{center}
        \begin{math}
        \epsfxsize#2cm
        \epsffile{#1}
        \end{math}
        \end{center}
        }
\newcommand{\vs}{\vspace{-0.25cm}}

\begin{document} 

\begin{center}
\large{\bf THE REACTION \begin{boldmath}$pp\to p \Lambda K^+$ \end{boldmath}
NEAR THRESHOLD}

\bigskip 

\bigskip

N. Kaiser\\

\bigskip

Physik Department T39, Technische Universit\"{a}t M\"{u}nchen,\\
    D-85747 Garching, Germany

\end{center}

\bigskip

\bigskip

\begin{abstract}
We analyze the recent total cross section data for $pp \to p\Lambda K^+$ near
threshold measured at COSY. Using an effective range approximation for the 
on-shell $p\Lambda$ S-wave final state interaction we extract from these data
the combination ${\cal K} = \sqrt{2|K_s|^2+|K_t|^2} = 0.38$ fm$^4$ of the
singlet ($K_s$) and triplet ($K_t$) threshold transition amplitudes. We present
an exploratory calculation of various (tree-level) vector and pseudoscalar 
meson exchange diagrams. Pointlike $\omega$-exchange alone and the combined
($\rho^0,\omega,K^{*+} $)-exchange can explain the experimental value of $\cal
K$. The pseudoscalar meson exchanges based on a SU(3) chiral Lagrangian turn 
out to be too large. However, when adding $\pi^0$-exchange in combination with
the resonant $\pi N  \to S_{11}(1650)\to K \Lambda$ transition and introducing
monopole form factors with a cut-off $\Lambda_c= 1.5$ GeV one is again
able to reproduce the experimental value of  $\cal K$. More exclusive
measurements are necessary to reveal the details of the $pp\to p\Lambda K^+$
production  mechanism. 
\end{abstract}


\bigskip

Accepted for publication in {\it Eur. Phys. J. A.}

\bigskip

\section{Introduction and summary}
With the advent of the proton cooler synchrotron COSY at J\"ulich high
precision data for associated strangeness production in proton-proton
collisions, $pp \to p \Lambda K^+$, have become available in the near threshold
region \cite{balew1,balew2,bilger}.  These data are of interest in several
respects. First, they can provide a possibility to test various theoretical
models of the strangeness dissociation mechanism for the nucleon. Secondly, the
cross  sections for strangeness production in the elementary $pp$-collision are
an  important input into transport model calculations of the strangeness
production in heavy ion collisions. The latter may provide information about
the hot and dense state of nuclear matter or the formation of the quark-gluon
plasma. Various dynamical models have been developed for the reaction $pp\to
p\Lambda K^+$ in refs.\cite{faeldt,sibir} which in particular focus on the role
of the $S_{11}(1650)$ nucleon resonance decaying into $K\Lambda $.   

In a recent work \cite{bkmcol} we have developed a novel approach to pion and 
eta production in proton-proton collisions, $pp\to pp\pi^0, pn \pi^+,pp \eta$, 
near threshold. In this approach one starts from the invariant T-matrix at 
threshold in the center-of-mass frame which is parametrized in terms of one 
(or two) constant threshold amplitudes. Close to threshold the relative
momentum of the nucleons in the final state is very small and their empirically
known strong S-wave interaction plays an essential role in the
description of the meson-production data. In fact is was found in 
ref.\cite{bkmcol} that in all three cases the energy dependence of the total 
cross section near threshold is completely and accurately determined by the 
three-body phase space and the on-shell S-wave NN final state interaction (in 
the case $pp\to pp\eta$ also the S-wave $\eta N$-interaction was included). 
Close to threshold the final state interaction can even be treated in effective
range approximation using the well-known values of the scattering lengths and 
effective range parameters. Note that ref.\cite{bkmcol} gives a (partial) 
derivation of such an approach to final state interaction in the context of
effective field theory (i.e. using only Feynman diagrams). Once one accepts 
such a phenomenological
separation of the (on-shell) final state interaction from the full production 
process, one can extract from the total cross section data an experimental 
value of the constant threshold amplitude parametrizing the T-matrix. In the 
next step a standard Feynman diagram calculation is performed for the 
center-of-mass T-matrix at threshold. It was stressed in ref.\cite{bkmcol} that
the evaluation of the Feynman diagrams has to be done fully relativistically 
since non-relativistic approximations will in general fail to reproduce 
correctly certain nucleon propagators. The source of this problem is the 
extreme kinematics of the meson-production process with the external 
center-of-mass momentum  $|\vec p\,| \simeq \sqrt{ M m_{\pi,\eta}}$ being 
proportional to the square root of the nucleon and meson mass. As a major 
result it was found in ref.\cite{bkmcol} that already the well-known 
tree-level (pseudoscalar and vector) meson exchange diagrams lead to
predictions for the constant threshold amplitudes which agree with the
corresponding experimental values within a few percent. It also turned out that
the short range ($\rho$ and $\omega$) vector meson exchange dominates over the 
long range ($\pi$ and $\eta$) pseudoscalar meson exchange for the processes
$pp\to pp \pi^0, pn\pi^+,pp\eta$ at threshold.  

The purpose of this work is to present a similar analysis for the kaon 
production channel $pp\to p\Lambda K^+$. After defining the threshold T-matrix
for $pp\to p\Lambda K^+$ in terms of a singlet ($K_s$) and a triplet ($K_t$) 
transition amplitude and implementing the $p\Lambda$ S-wave final state 
interaction in effective range approximation, we extract from the COSY
data an experimental value for the combination ${\cal K}=\sqrt{2|K_s|^2+|K_t|^2
}= 0.38$ fm$^4$. Next we perform a relativistic Feynman diagram calculation of
various vector meson ($\rho^0, \omega, K^{*+}$) and pseudoscalar meson ($\pi^0,
\eta, K^+$) exchange diagrams with vertices given by a SU(3) symmetric (chiral)
meson-baryon Lagrangian. We first evaluate these tree diagrams 
straightforwardly from the relativistic SU(3) Lagrangian not introducing  
ad hoc meson-nucleon form factors. The latter are a model-dependent and 
unobservable concept to account in some sense for the finite size of the
hadrons involved. It is found that pointlike $\omega$-exchange alone with
coupling constants given by SU(3) symmetry  can well reproduce the experimental
value of ${\cal K}=0.38$ fm$^4$. The total ($\rho^0,\omega,K^{*+}$) vector 
meson exchange leads to ${\cal K}=0.45$ fm$^4$ which is about 20\% too large. 
Taking into account that possible SU(3) breaking effects and the uncertainty 
of the vector meson baryon coupling constant are of similar size, one can 
argue that the (pointlike) total vector meson exchange is still capable to 
explain the experimental value ${\cal K}=0.38$ fm$^4$. The pseudoscalar meson
($\pi^0, \eta, K^+)$ exchange diagrams with vertices given by 
the next-to-leading order SU(3) chiral Lagrangian give rise to rather large  
individual contributions. There is a tendency for cancelation between different
types of diagrams, but with the SU(3) chiral meson-baryon vertices the effect 
is not pronounced enough. Interestingly, the (pointlike) $K^+$-exchange alone
leads to ${\cal K}=0.41$ fm$^4$ and if one omits the $\pi^0$-exchange one 
obtains ${\cal K}=0.43$ fm$^4$ from the remaining ($\rho^0, \omega, K^{*+},
K^+, \eta$)-exchange diagrams. It therefore seems that the transition amplitude
for $\pi N \to K \Lambda$ as given by the SU(3) chiral Lagrangian is  not
realistic. This is also underlined by  the  work of ref.\cite{kww} in which
these chiral  amplitudes have been iterated to infinite orders via a coupled
channel Lippmann-Schwinger equation and this way a good simultaneous fit of all
available low energy data for  pion (and photon) induced $(\eta, K)$-production
could be found. We follow here refs.\cite{faeldt,sibir} and add a 
$\pi^0$-exchange diagram involving the resonant $\pi N \to S_{11}(1650)\to 
K \Lambda $ transition. We assume the transition strength to be such that this
process alone reproduces the near threshold data for $pp\to p\Lambda K^+$. If 
we furthermore introduce at each meson-baryon vertex a monopole form factor 
with a cutoff $\Lambda_c =1.5$ GeV  we finally end up with a total sum of the 
$S_{11}(1650)$-excitation graph and the vector and pseudoscalar meson exchange 
diagrams which again is in good agreement with the experimental value 
${\cal K}=0.38$ fm$^4$.

It becomes also clear from our analysis that the unpolarized total cross 
section data for $pp \to p\Lambda K^+$ do not provide enough information to 
distinguish different production mechanisms. In essence the information given 
by the total cross section data can be condensed into a single number, namely
${\cal K} = \sqrt{2|K_s|^2+|K_t|^2} = 0.38$ fm$^4$. As we have demonstrated
here, one can find various subprocesses (as e.g. the vector meson exchange,
the $K^+$-exchange or the $S_{11}(1650)$-excitation focussed on in
ref.\cite{faeldt}) which alone can reproduce this value. More exclusive 
measurements of angular distributions and polarization observables for which
one does not average out the kinematical complexity of the process $pp\to
p\Lambda K^+$ due to spin and 
the three-particle final state are needed. Such measurements have been started 
\cite{bilger} and we expect more data to come from COSY in the near future. 

\section{Threshold T-Matrix}
The T-matrix for kaon and lambda-hyperon production in proton-proton
collisions, $p_1(\vec p\,) +p_2(-\vec p\,) \to p +\Lambda +K^+$, at threshold 
in the center-of-mass frame reads, 
\begin{equation} T^{\rm cm}_{\rm th}(pp\to p\Lambda K^+) = {K_s\over \sqrt 3}
\, (i \, \vec\sigma_1 - i\, \vec \sigma_2+ \vec \sigma_1 \times \vec
\sigma_2)\cdot \vec p + {K_t\over \sqrt 3} \, i(\vec \sigma_1 + \vec \sigma_2)
\cdot \vec p \,\,, \end{equation}
where $\vec p$ is the proton center-of-mass momentum with $|\vec p\,| = 861.5$
MeV at threshold. The spin-operator $\vec \sigma_1$ is understood to be
sandwiched between the spin-states of the ingoing proton $p_1(\vec p\,)$ and 
the outgoing proton, while $\vec \sigma_2$ acts between the proton $p_2(-\vec 
p\,)$ and the outgoing lambda. The (complex) amplitude $K_s$ belongs to the
singlet transition $^3P_0 \to $ $^1S_0 s$ and the amplitude $K_t$ belongs to
the triplet transition $^3P_1\to $ $^3S_1 s$. The factor $1/\sqrt3$ was taken
out for convenience since it appears naturally in a calculation employing SU(3)
symmetry. We follow now the successful approach to pion and eta production  of 
ref.\cite{bkmcol} and 
assume the T-matrix to be constant in the near threshold region and the energy 
dependence of the total cross section to be given by the three-body phase 
space and the (on-shell) $p\Lambda$ S-wave final state interaction. Since the 
outgoing proton and lambda have small relative momentum one treat their S-wave
interaction in effective range approximation, i.e. in terms of the 
singlet and triplet $p\Lambda$ scattering lengths and the singlet and triplet
$p\Lambda$ effective range parameters. Experimental evidence \cite{plevid} and 
model calculations \cite{plmodel} suggest that these are rather similar for the
$^1S_0$ and $^3S_1$ $p\Lambda$-states. In this case the unpolarized total cross
section for $ pp\to p \Lambda K^+$ including the  $p\Lambda$ S-wave final state
interaction reads,    
\begin{eqnarray} \sigma_{\rm tot}(\epsilon) &=& (2|K_s|^2+|K_t|^2)  \,{M^3
M_\Lambda \sqrt{s-4M^2} \over 48\pi^3 s^{3/2}}  \nonumber \\ &&  
\times \int_{M+M_\Lambda}^{\sqrt s -m_K} {d W \over W}  \sqrt{\lambda(W^2,M^2,
M_\Lambda^2)\,\lambda(W^2,m_K^2,s)}\, \bar F_{p\Lambda}(W) \,\,, \end{eqnarray}
with $ \epsilon = \sqrt s -M-M_\Lambda-m_K$ the center-of-mass excess energy.  
$M, M_\Lambda$ and $m_K$ denote the proton, lambda and (charged) kaon mass. 
$W$ is the $p\Lambda$ invariant mass with values between $M+M_\Lambda$ and the
kinematical endpoint $\sqrt s - m_K$, and $\lambda(x,y,z) =x^2+y^2+z^2-2yz -2xz
-2xy$ denotes the K\"allen or triangle function. The correction factor from the
(equal singlet and triplet) $p\Lambda$ S-wave final state interaction reads in
effective range approximation,
\begin{equation} \bar F_{p\Lambda}(W) = \bigg[1 + {\bar a (\bar a +\bar r) 
\over 4W^2} \lambda(W^2,M^2,M_\Lambda^2) + {\bar a^2 \bar r^2 \over 64 W^4
}\lambda^2(W^2, M^2, M_\Lambda^2) \bigg]^{-1} \,\,.\end{equation}
We use for the $p\Lambda$ scattering length and effective range parameter the
values $\bar a = 2.0$ fm and $\bar r =1.0$ fm as extracted in ref.\cite{balew3}
from the Dalitz plot distributions of the $pp\to p \Lambda K^+$ data and the
low energy  elastic $p\Lambda$ scattering cross sections. Using eqs.(2,3) for 
the total cross section, a best fit of the seven  COSY data points near
threshold \cite{balew2} leads to the following experimental value of the 
combination
\begin{equation} {\cal K} = \sqrt{2|K_s|^2+|K_t|^2} = 0.38\, {\rm fm}^4
\,\,. \end{equation}\vs\vs\vs
\begin{table}[hbt]
\begin{center}
\begin{tabular}{|c|ccccccc|}
    \hline
    $\epsilon$~[MeV]&0.68&1.68&2.68&3.68&4.68&5.68&6.68 \\
    \hline
    $\sigma_{\rm tot}^{\rm exp}$~[nb]&$2.1\pm 0.2$&$13.4 \pm 0.7$&$36.6 \pm 2.6
$&$63.0\pm 3.1$&$92.2 \pm 6.5$&$135. \pm 11.$&$164. \pm 10.$ \\
    \hline
  $\sigma_{\rm tot}^{\rm fit}$~[nb]&2.5&14.2&34.0&60.5&92.7&130.&171.\\
    \hline
  \end{tabular}
\end{center}
{\it Tab.1: Total cross sections for $pp\to p\Lambda K^+$. The 
data are taken from ref.\cite{balew2} and the fit is described in the text.}
\end{table}
\bild{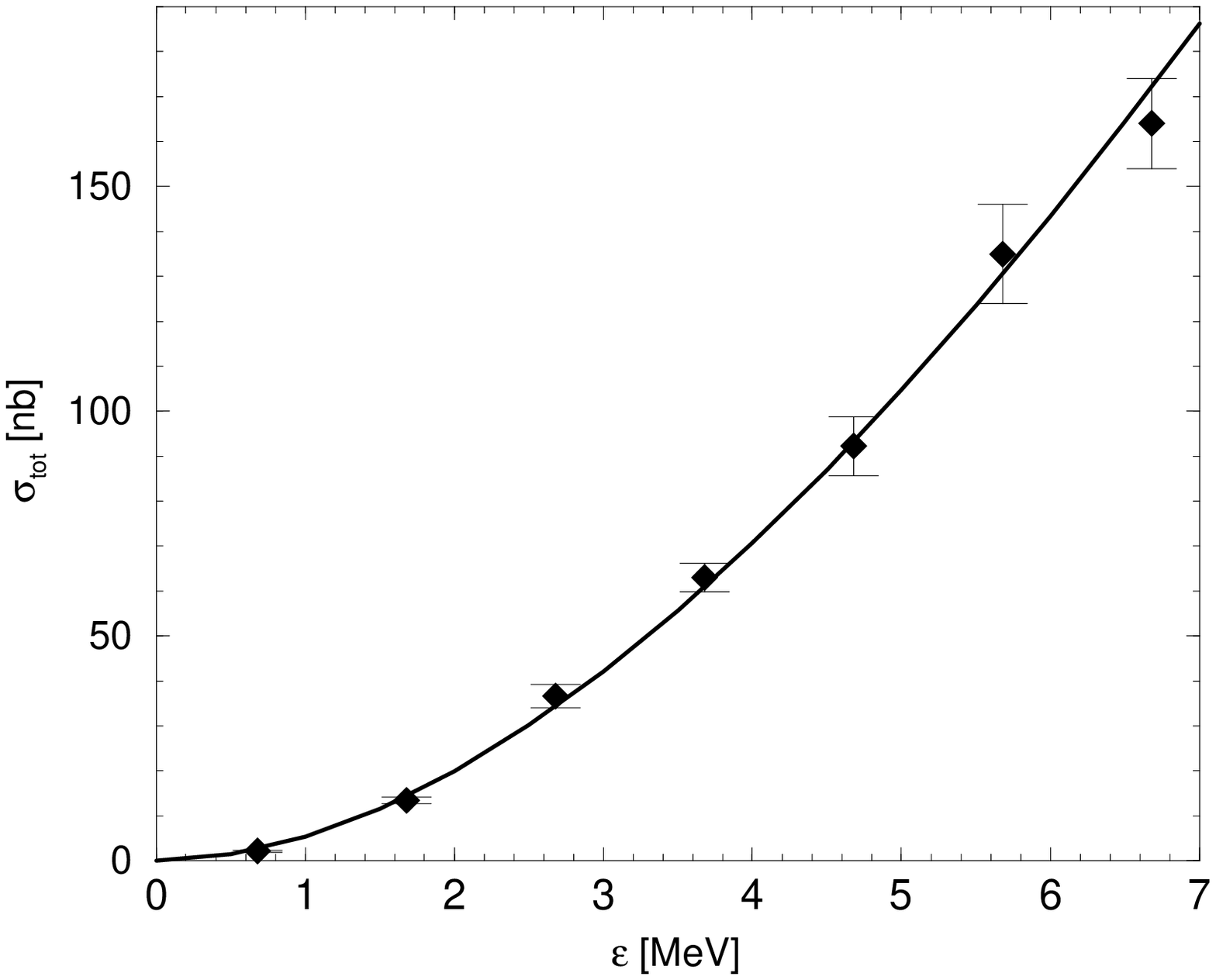}{14}
{\it Fig.1: Total cross sections for $pp\to p\Lambda K^+$ as a function of the
center-of-mass excess energy $\epsilon = \sqrt s - M-M_\Lambda - m_K$. The 
data are taken from ref.\cite{balew2} and the full line is calculated with
${\cal K}= 0.38$ fm$^4$ and $p\Lambda$ final state interaction.}
\bigskip

The resulting  fit values of $\sigma_{\rm tot}$ are given in Tab.1 and the
energy dependent total cross section is shown in Fig.1 for excess energies
$\epsilon\leq 7$ MeV. Note that the best fit values at the two lowest energies 
$\epsilon = 0.68$ MeV and $\epsilon= 1.68$ MeV lie somewhat outside the 
experimental error band. This may be due to the neglect of the $pK^+$ Coulomb
interaction. A similar slight overestimation of the data points closest to 
threshold was also observed in ref.\cite{bkmcol} for the reaction $pp \to pp
\pi^0$. Compared to the processes $pp\to pp\pi^0,pn\pi^+,pp\eta$ studied
in ref.\cite{bkmcol} the $p\Lambda$ final state interaction plays here a much 
less important role. This is also visible from Fig.1 which shows that the data 
follow already approximately the pure three-body phase space behavior, i.e. $
\sigma_{\rm  tot}(\epsilon) \sim \epsilon^2$. Recently, three additional data
points near threshold have been measured at COSY \cite{eyrich}: $\sigma_{\rm
tot}(\epsilon=8.6\,{\rm MeV}) = (344\pm41)$\,nb,  $\sigma_{\rm tot}(\epsilon=
10.9\,{\rm MeV}) = (385\pm 27)$\,nb and  $\sigma_{\rm tot}(\epsilon=13.2\,{\rm 
MeV}) = (505\pm 33)$\,nb. These values are well reproduced by the present fit
which predicts: $261$\,nb, $383$\,nb and $518$\,nb, respectively. We note aside
that the two data points at higher excess energies $\sigma_{\rm tot}(\epsilon=
55\,{\rm MeV})=(2.7\pm0.3) \,\mu$b and $\sigma_{\rm tot}(\epsilon= 138\, {\rm 
MeV})=(12.0\pm0.4)\, \mu$b \cite{bilger} come out according to eq.(2) as $3.9
\,\mu$b and $11.5\,\mu$b. The good agreement for $\epsilon = 138$ MeV should be
regarded as accidental, since the same experiment \cite{bilger} has measured a
negative $\Lambda$ recoil polarization, which can only result from S- and
P-wave  interference terms. 

\section{Diagrammatic approach}
In this section we will present an exploratory calculation of various vector
and pseudoscalar meson exchange diagrams contributing to the threshold
amplitudes $K_{s,t}$ defined in eq.(1). The coupling of vector mesons to
baryons is described by the SU(3) symmetric Lagrangian,
\begin{equation} {\cal L}_{VB} = {g_V\over 2}\,\Big\{ {\rm tr}(\bar B\gamma_\mu
[ V^\mu,B]) + {\rm tr}(\bar B\gamma_\mu B) {\rm tr}( V^\mu) \Big\}\,\,. 
\end{equation}
The SU(3)-matrices $B$ and $V^\mu$ collect the octet baryon fields ($N,\Lambda,
\Sigma,\Xi$) and the octet vector meson fields ($\rho, \omega, K^*, \phi$),
respectively. The form eq.(5) implies the relations $g_{\rho N} = g_V/2 = 
g_{\omega N}/3$ and $g_{\phi N}=0$, and we use $g_V \simeq 6$ for the vector
meson coupling constant. The pseudoscalar meson-baryon interaction is given by
the chiral Lagrangians, 
\begin{equation} {\cal L}^{(1)}_{\phi B} = {i\over 8f_\pi^2} {\rm tr}(\bar B
\gamma_\mu [[\phi,\partial^\mu \phi],B]) +{D\over 2f_\pi} {\rm tr}(\bar B
\gamma_5 \gamma_\mu \{\partial^\mu \phi,B\}) +{F\over 2f_\pi} {\rm tr}(\bar B
\gamma_5 \gamma_\mu [\partial^\mu \phi,B]) \,\,,\end{equation}
\begin{equation} {\cal L}^{(2)}_{\phi B} = b_D {\rm tr}(\bar B\{\chi_+,B\})+ 
b_F {\rm tr}(\bar B[\chi_+,B])+ b_0 {\rm tr}(\bar BB){\rm tr}(\chi_+)\,, \quad
 \chi_+= 2 \chi_0 - {1\over 4f_\pi^2} \{\phi,\{ \phi, \chi_0\}\} \,,
\end{equation}
with the diagonal matrix $\chi_0 ={\rm diag}(m_\pi^2,m_\pi^2,2m_K^2-m_\pi^2)$. 
The SU(3)-matrix $\phi$ collects the octet pseudoscalar meson fields ($\pi, K,
\eta$) and $f_\pi = 92.4$ MeV is the weak pion decay constant. $D\simeq 0.75$
and $F\simeq 0.50$ are the axial vector coupling constants as determined from
semi-leptonic hyperon decays. The here most relevant $N\Lambda K$ coupling
constant is given according to eq.(6) as $g_{N\Lambda K}=-(D+3F)(M+M_\Lambda)/
(2\sqrt 3 f_\pi) = - 14.4$, which is rather close to the empirical value 
$g_{N\Lambda K}= - 13.2$ of ref.\cite{martin}. For the second order chiral
Lagrangian ${\cal L}^{(2)}_{\phi B}$ only the explicit chiral symmetry breaking
terms (linear in the quark masses) are displayed in eq.(7). The coefficients
$b_{D,F,0}$ are related to mass splittings in the baryon octet, the $\pi N$
$\sigma$-term $\sigma_N(0) = (45\pm 8)$ MeV and the (scalar) strangeness
content of the nucleon. We use the values $b_D= 0.066$ GeV$^{-1}$, $b_F =
-0.213$ GeV$^{-1}$ and $b_0 = -0.304$ GeV$^{-1}$, found in ref.\cite{kww}. At
second order there exists in addition a large set of double-derivative terms
($\sim \partial_\mu \phi \partial_\nu \phi$)  with a priori unknown
coefficients. For the process $NN\to NN\pi$ it was observed  in 
ref.\cite{bkmcol} that the explicit chiral symmetry breaking term is dominant
at second order and therefore we neglect here the double-derivative terms with
unknown coefficients.   

\subsection{Vector meson exchange}
The vector meson exchange diagrams contributing to $pp\to p\Lambda K^+$ are 
shown in Fig.2. Note that according to eq.(5) both the $\Lambda\Lambda \rho^0$ 
and $\Lambda \Sigma^0 \rho^0$ coupling vanish and therefore no $\rho^0$ 
exchange occurs in the right hand diagram where the $K^+$-meson is emitted 
from  the  proton line before the vector meson exchange.

\begin{center}
\SetScale{0.8}
\SetWidth{1.5}
  \begin{picture}(364,72)

\Line(75,0)(75,85)
\Line(142,0)(142,85)
\DashLine(75,30)(142,30){7}
\DashLine(142,50)(182,85){7}
\Text(85,35)[]{$\omega, \rho^0$}

\Line(290,0)(290,85)
\Line(356,0)(356,85)
\DashLine(290,55)(356,55){7}
\DashLine(356,32)(400,70){7}
\Text(260,57)[]{$\omega, K^{*+}$}
\end{picture}

\medskip

{\it Fig.2: Vector meson exchange diagrams contributing to $pp\to p\Lambda
K^+$.}
\end{center}

Straightforward evaluation at threshold of the diagrams shown in Fig.2 gives
for $\omega$-exchange, 
\begin{eqnarray} K_t^{(\omega)} &=& {3g_V^2 (D+3F)m_K(4M-m') \over 8M f_\pi 
(m_\omega^2 + M m') (2M+m')m'} = 0.38 \, {\rm fm}^4 \,\,, \nonumber \\
K_s^{(\omega)} &=& {3g_V^2 (D+3F)m_K(m'-M) \over 4Mf_\pi (m_\omega^2 +  M m')
(2M+m')m'} = -0.07 \,  {\rm fm}^4 \,\,,  \end{eqnarray}
and for $\rho^0$-exchange,
\begin{eqnarray} K_t^{(\rho)} &=& {g_V^2 (D+3F)m_K(2+\kappa_\rho) \over 16M f_\pi (m_\rho^2 + M m') (2M+m')}\bigg[{m'\over 4M} \kappa_\rho-1 \bigg] = 0.01 \, 
{\rm fm}^4 \,\,, \nonumber \\ K_s^{(\rho)} &=& {g_V^2 (D+3F)m_K \over 16 M 
f_\pi (m_\rho^2 +  M m') (2M+m')} \bigg[ 2 +{3\over 2} \kappa_\rho - {m'\over
8M} \kappa_\rho ( 6+5\kappa_\rho) \bigg] = -0.12 \, {\rm fm}^4
\,\,. \end{eqnarray} 
Note that we have included the large anomalous tensor-to-vector coupling ratio
$\kappa_\rho \simeq 6$ in the $pp\rho^0$-vertex. For the $pp\omega$-vertex the 
tensor coupling is known to be small $\kappa_\omega\simeq 0$. The abbreviation
$m'$ stands for $m'=M_\Lambda-M+m_K=671.0$ MeV. Furthermore, one finds from
$K^{*+}$-exchange,
\begin{equation} K_t^{(K^*)} =2 K_s^{(K^*)} = -{3g_V^2 F m_K \over 2M f_\pi 
(m_{K^*}^2 +M m') m'} = -0.24 \, {\rm fm}^4 \,\,.  \end{equation}
For the $K^{*+}$-exchange diagram the intermediate state baryon can be either 
a $\Lambda$ or a $\Sigma^0$ and we neglected the small mass difference
$M_{\Sigma^0}-M_\Lambda= 77$ MeV in eq.(10). In the absence of any empirical
evidence, we did not include an anomalous tensor coupling in the $p\Lambda
K^{*+}$-vertex.  One observes that $\omega$-meson exchange alone, eq.(8), with
${\cal K}^{(\omega)} = 0.39$ fm$^4$, well reproduces the empirical value ${\cal
K}=0.38$ fm$^4$ extracted from the COSY data. Summing up all vector meson
($\omega,\rho^0,K^{*+}$)-exchange contributions one gets ${\cal K}^{(V)}= 0.45$
fm$^4$ which is about 20\% too large. Taking into account, that the uncertainty
of $g_V$ and possible SU(3) breaking effects are of similar size one can still
argue that (pointlike) vector meson exchange is able to explain the
experimental value ${\cal K}=0.38$ fm$^4$. 

\subsection{Pseudoscalar meson exchange} 
The pseudoscalar meson exchange diagrams contributing to $pp\to p\Lambda K^+$
are shown in Fig.3. Two types of diagrams are possible, one where the meson 
rescatters via a chiral contact vertex, and another one with a baryon 
propagating in the intermediate state.  
\begin{center}
\SetScale{0.8}
\SetWidth{1.5}
  \begin{picture}(364,72)
\Line(10,0)(10,85)
\Line(80,0)(80,85)
\DashLine(10,42.5)(80,42.5){7}
\DashLine(80,42.5)(120,85){7}
\Text(36,45)[]{$\pi^0,K^+,\eta$}

\Line(175,0)(175,85)
\Line(246,0)(246,85)
\DashLine(175,30)(246,30){7}
\DashLine(246,50)(290,85){7}
\Text(167,33)[]{$\pi^0,\eta$}

\Line(350,0)(350,85)
\Line(420,0)(420,85)
\DashLine(350,55)(420,55){7}
\DashLine(420,28)(460,80){7}
\Text(309,35)[]{$\pi^0,K^+,\eta$}
  \end{picture}

\medskip

{\it Fig.3: Pseudoscalar meson exchange diagrams contributing to $pp\to
p\Lambda K^+$.}

\end{center}

The calculation of the rescattering type diagrams in Fig.3 gives for 
$\pi^0$-exchange, 
\begin{equation} K_t^{(\pi)} = 2K_s^{(\pi)} = {D+F \over 16 f_\pi^3 (m_\pi^2 +
M m') } \Big[4(b_D+3b_F)(m_K^2+m_\pi^2) -3(m'+m_K) \Big] = -0.95 \, {\rm fm}^4
\,\,, \end{equation}
for $K^+$-exchange,
\begin{equation} K_t^{(K)} = -2K_s^{(K)} = -{D+3F \over 4f_\pi^3 (m_K^2 + M m')
} \Big[ 8(b_0+b_D)m_K^2+m'+m_K \Big] = -0.87 \, {\rm fm}^4 \,\,,
\end{equation}
and for $\eta$-exchange,
\begin{equation} K_t^{(\eta)} = 2K_s^{(\eta)} = {D-3F \over 48 f_\pi^3
(m_\eta^2 + M m') } \Big[ 4(b_D+3b_F)(5m_K^2-3 m_\pi^2) +9(m'+m_K) \Big] =
-0.25 \, {\rm fm}^4 \,\,. \end{equation}
From the other diagrams in Fig.3 with a baryon propagating in the intermediate 
state one finds for $\pi^0$-exchange,
\begin{equation} K_t^{(\pi)} = 2K_s^{(\pi)} = {(D+F) m_K \over 8f_\pi^3(m_\pi^2
+ M m')(2M+m') }\Big[ 4D(D-F)M+(3D^2+2DF+3F^2)m'\Big]  = 0.26 \, {\rm fm}^4
\,\,, \end{equation}
for $K^+$-exchange, 
\begin{equation} K_t^{(K)} = -2K_s^{(K)} = {(D+3F)(D^2+3F^2) m_K \over6f_\pi^3
(m_K^2 + M  m') }  = 0.54 \, {\rm fm}^4 \,\,, \end{equation}
and for $\eta$-exchange,
\begin{equation} K_t^{(\eta)} = 2K_s^{(\eta)} = {(9F^2-D^2) m_K \over 24f_\pi^3
(m_\eta^2 +M m') (2M+m')} \Big[ D(4M+m')+3 F m'\Big] = 0.12 \, {\rm fm}^4 \,\,.
\end{equation}
Numerically, the pseudoscalar meson exchange contributions generated by the
SU(3) chiral Lagrangians eqs.(6,7) are too large, $K_t^{(ps)} =- 1.16$
fm$^4$, $K_s^{(ps)} =- 0.25$ fm$^4$. When combined with the vector meson
exchange terms one would get ${\cal K}^{(V+ps)}= 1.28$ fm$^4$, which is about a
factor 3 too large. Note that there is some tendency for cancelation between
the rescattering type diagrams and the other ones, but the effect is not 
pronounced enough. It is interesting to observe that the pointlike 
$K^+$-exchange alone eqs.(12,15) gives ${\cal K}^{(K)}= 0.41 $ fm$^4$ and if
one omits the $\pi^0$-exchange eqs.(11,14) one obtains from the remaining
contributions ${\cal K}^{(V+K+\eta)} =0.43$ fm$^4$. Both these numbers are 
close to the experimental value ${\cal K}=0.38$
fm$^4$. It therefore seems that the tree level $\pi N \to K\Lambda$ transition
amplitude as given by the next-to-leading order chiral Lagrangian eqs.(5,6,7)
has some unrealistic features. This is also underlined by the recent work of
ref.\cite{kww} in which these chiral amplitudes have been iterated to infinite
order via a Lippmann-Schwinger equation and this way a good simultaneous 
fit of all available low energy data for pion (and photon) induced $(\eta,
K)$-production could be found. Instead of performing a similar coupled channel 
calculation we follow here refs.\cite{faeldt,sibir} and add a further 
$\pi^0$-exchange diagram involving the resonant $\pi N \to S_{11}(1650)\to
K\Lambda $ transition. As argued in ref.\cite{faeldt} we assume the 
corresponding transition strength to be so large that this process alone 
can reproduce the near threshold data for $pp\to p \Lambda K^+$. In this case 
one has the following contribution from $S_{11}(1650)$-excitation to the
triplet and  singlet threshold amplitudes, 
\begin{equation} K_t^{(N^*)} = 2 K_s^{(N^*)} = 0.31 \, {\rm fm}^4  \,\,. 
\end{equation}       
We have convinced ourselves that the direct evaluation of the
$S_{11}(1650)$-excitation graph leads to values of similar size taking into
account  the empirical ranges of the $S_{11}(1650)$-resonance mass and partial
decay widths into $\pi N$ and $K\Lambda$. Let us furthermore introduce a
monopole form factor at each (off-shell) meson-baryon vertex with a cut-off 
$\Lambda_c =1.5$ GeV. Such values of the cut-off $\Lambda_c$ are typically used
in one-boson exchange models \cite{mach} of the NN-interaction for both the 
pion- and vector-meson-nucleon vertex. This modification of the diagrams 
brings a reduction factor $(1+Mm'/\Lambda_c^2)^{-2}$  for the vector and
pseudoscalar meson exchange graphs. Note that in the case of the 
$S_{11}(1650)$-resonance contribution the form factor effect is already 
included in the number given in eq.(17). Summing up all the contributions due 
to $S_{11}(1650))$-excitation as well pseudoscalar and vector meson exchange
(including the form factor) one finds $K_t^{(\rm tot)}= -0.31$ fm$^4$ and
$K_s^{(\rm tot)}=-0.18$ fm$^4$. The resulting value ${\cal K}^{(\rm tot)}=0.40
$ fm$^4$ is again in good agreement with the empirical value ${\cal K}=0.38$ 
fm$^4$. 

Evidently, the main lesson to be learned from the present exploratory 
calculation is that there are in fact various subprocesses (like pointlike 
vector meson or $K^+$-exchange or the $S_{11}(1650)$-resonance excitation)
which alone can explain the near threshold data for $pp\to p\Lambda K^+$. With
reasonable assumptions on the various coupling strengths and the cut-off
$\Lambda_c$ entering the (unobservable) meson-nucleon form factor one finds 
that also the total sum of many processes is able to reproduce the near 
threshold data for $pp\to p\Lambda K^+$. This is possible because of 
cancelations between terms of different sign and because of the freedom the 
to shift strength between the singlet ($K_s$) and triplet ($K_t$) threshold 
amplitude. Only more exclusive data (like angular distributions and
polarization observables) can help to
distinguish different $pp\to p\Lambda K^+$ production mechanisms.

Finally, we like to comment on the recently measured process $pp\to p\Sigma^0
K^+$. It was found experimentally \cite{eyrich} that the corresponding total 
cross sections near threshold are about a factor 30 smaller than those for
$pp\to p\Lambda K^+$ (at equivalent excess energies). $\omega$-meson exchange 
or $K^+$-exchange might offer an explanation for this suppression via the small
SU(3)-ratio $g_{N\Sigma K}/g_{N\Lambda K}=-\sqrt{3}/9 \simeq -0.19$. An
explicit calculation of the $\omega$-exchange and $K^+$-exchange (using now
$m'=M_{\Sigma^0}-M+m_K=748$ MeV in eqs.(8,12,15)) gives for the $p\Sigma^0
K^+$-channel values of ${\cal K}_{\Sigma^0}^{(\omega)}=0.06$ fm$^4$ and ${\cal
K}_{\Sigma^0}^{(K)}=0.09$ fm$^4$, i.e. a suppression factor of about 40 or 20  
for the total cross  section. However, such considerations may be too
simplistic in the light of possible strong coupled channel effects \cite{kww}
between $K\Lambda$- and $K\Sigma$-states.  

\end{document}